\documentclass[twocolumn, times]{aastex631} 
\usepackage{graphicx}
\usepackage{booktabs}
\usepackage{hyperref}
\usepackage{amsmath}

\newcommand{\oiii}{[O\,\textsc{iii}]~}
\newcommand{\sii}{[S\,\textsc{ii}]~}

\newcommand{\orl}{O\,\textsc{ii}~}
\newcommand{\hii}{H\,\textsc{ii}~}

\newcommand{\hb}{H$\beta$~}

\begin{document}

\title{SDSS-V LVM: Mapping of plasma temperatures in the Lagoon Nebula reveals the origin \\
of long-standing chemical abundance discrepancies}


\author[0009-0000-3962-103X]{Amrita Singh} 
\newcommand{\UChile}{\affiliation{Departamento de Astronom\'{i}a, Universidad de Chile, Camino del Observatorio 1515, Las Condes, Santiago, Chile}} 
\UChile

\author[0000-0003-4218-3944]{Guillermo A. Blanc}
\affiliation{Observatories of the Carnegie Institution for Science, 813 Santa Barbara Street, Pasadena, CA 91101, USA}
\UChile


\author[0000-0001-6551-3091]{Kathryn Kreckel$^{\dagger}$}
\newcommand{\heidelberg}{\affiliation{Astronomisches Rechen-Institut, Zentrum f\"{u}r Astronomie der Universit\"{a}t Heidelberg, M\"{o}nchhofstra\ss e 12-14, D-69120 Heidelberg, Germany}}
\heidelberg

\author[0000-0002-6972-6411]{J. E. M\'endez-Delgado$^{\dagger}$}
\affiliation{Instituto de Astronom\'ia, Universidad Nacional Aut\'onoma de M\'exico, Ap. 70-264, 04510 CDMX, M\'exico}

\author[0000-0001-5801-6724]{Christophe Morisset$^{\dagger}$}
\affiliation{Instituto de Astronomía (IA), Universidad Nacional Autónoma de México, Apdo. Postal 106, C.P., 22800 Ensenada, Baja California, México}
\affiliation{Instituto de Ciencias Físicas, Universidad Nacional Autónoma de México, Av. Universidad s/n, 62210 Cuernavaca, Morelos, México}

\altaffiliation{$^{\dagger}$ These three authors contributed equally to this work and are listed alphabetically.}

\author[0000-0001-6444-9307]{Sebastián F.\ Sánchez}
\affiliation{Instituto de Astronom\'ia, Universidad Nacional Auton\'oma de M\'exico, A.P. 106, Ensenada 22800, BC, Mexico}
\affiliation{Instituto de Astrof\'\i sica de Canarias, La Laguna, Tenerife, E-38200, Spain}

\author[0000-0002-4755-118X]{Oleg Egorov}
\heidelberg

\author[0000-0002-7339-3170]{Niv Drory}
\affiliation{McDonald Observatory, The University of Texas at Austin, 1 University Station, Austin, TX 78712-0259, USA}

\author[0000-0002-5320-2568]{Nimisha Kumari}
\newcommand{\stsci}{\affiliation{Space Telescope Science Institute, 3700 San Martin Drive, Baltimore, MD 21218, USA}}
\stsci

\author[0000-0003-2496-1247]{Fu-Heng Liang}
\heidelberg

\author[0000-0002-8549-4083]{Enrico Congiu}
\affiliation{European Southern Observatory, Alonso de C\'ordova 3107, Casilla 19, Santiago 19001, Chile}

\author[0000-0002-8931-2398]{Alfredo Mej\'ia-Narv\'aez}
\UChile

\author[0000-0002-2368-6469]{Evelyn J.\ Johnston}
\affiliation{Instituto de Estudios Astrof\'isicos, Facultad de Ingenier\'ia y Ciencias, Universidad Diego Portales, Av. Ej\'ercito Libertador 441, Santiago, Chile}

\author[0000-0003-3526-5052]{Jos\'e G. Fern\'andez-Trincado}
\affiliation{Centro de investigaci\'on en Astronom\'ia, Facultad de Ingenier\'ia, Ciencia y Tecnolog\'ia, Universidad Bernardo O'Higgins, Av. Viel 1497, Santiago, 8370993, Chile}

\author[0009-0007-4271-6444]{Hemanth Bommireddy}
\UChile

\author[0000-0002-0635-2264]{Abhinna Sundar Samantaray}
\heidelberg

\author[0000-0001-7641-5235]{Thomas Hilder}
\affiliation{School of Physics and Astronomy, Monash University VIC 3800, Australia}

\author[0009-0009-0081-4323]{Rodolfo de J. Zermeño}
\affiliation{Instituto de Astronom\'ia, Universidad Nacional Aut\'onoma de M\'exico, Ap. 70-264, 04510 CDMX, M\'exico}

\author[0000-0002-6425-6879]{Ivan Katkov}
\affiliation{New York University Abu Dhabi, PO Box 129188, Abu Dhabi, UAE}

\author[0000-0002-7955-7359]{Dhanesh Krishnarao}
\affiliation{Department of Physics, Colorado College, 14 East Cache la Poudre St., Colorado Springs, CO, 80903, USA}

\author[0000-0001-9852-1610]{Juna A.\  Kollmeier}
\affiliation{Canadian Institute for Theoretical Astrophysics, University of Toronto, Toronto, ON M5S-98H, Canada}
\affiliation{Observatories of the Carnegie Institution for Science, 813 Santa Barbara Street, Pasadena, CA 91101, USA}

\correspondingauthor{Amrita Singh}
\email{amrita@das.uchile.cl}

\begin{abstract}

Knowing the chemical composition of interstellar gas is essential for understanding the chemical evolution of galaxies. Ionized nebulae provide key laboratories for this purpose, where chemical abundances are commonly derived from emission lines of heavy ions such as doubly ionized oxygen, O$^{++}$. Abundances can be inferred either from  collisionally excited lines (CELs) or from recombination lines (RLs). However, for more than 80 years, the former have yielded systematically lower abundances than the latter. We map this “abundance discrepancy” using the first spatially resolved integral field unit (IFU) observations of O$^{++}$ RLs and CELs across a Milky Way \hii region: the Lagoon Nebula (a.k.a. M8). Modeling the data using a plasma temperature perturbation model, we present unequivocal evidence that electron temperature ($T_e$) fluctuations drive the abundance discrepancy in M8. We find a tight correlation between the abundance discrepancy and the difference between $T_e$ derived from the classical CEL method and a hybrid RL/CEL approach. We identify this $T_e$ discrepancy as an observational proxy for the amplitude of the thermal fluctuations, providing a new, direct way to quantify their effect on nebular abundances. The correlation constrains the mean electron temperature of the gas. Adopting this mean temperature reconciles CEL and RL abundances, effectively eliminating the abundance discrepancy.

\end{abstract}


\section{Introduction}
\label{sec:sec1}

The chemical composition of galaxies reflect the cumulative history of star formation, stellar nucleosynthesis, astration, supernova explosions, gas and dust chemical redistribution and depletion processes, and the inflow, mixing, and dispersal of gas in and out of galaxies across cosmic time. These processes underpin a wide range of phenomena in astrophysics, making chemical abundances a key parameter in stellar and galaxy evolution models \citep{osterbrock06, matteucci12, nomoto13, kewley19, blanc19, kobayashi20}.

In star-forming galaxies, \hii regions mark sites of recent star formation, where radiation from young massive stars ionizes the surrounding neutral gas. The resulting plasma emits a spectrum rich in emission lines from multiple elements, whose relative intensities provide primary diagnostics of the physical conditions -- electron density ($n_e$), and electron temperature ($T_e$) -- and chemical composition of the ionized gas.

The abundance of heavy elements in the interstellar medium (ISM), measured with respect to that of hydrogen (H), provides a fundamental tracer of galaxy evolution. Oxygen, the most abundant heavy element, is widely used as a proxy for measuring the overall heavy element content of the ISM  \citep[e.g.][]{kewley19, blanc19, maiolino19}. The strong collisionally excited lines (CELs) of O$^{++}$ are readily detected in the optical spectra of \hii regions. Faint auroral CELs, historically hard to detect in many galactic and nearby \hii regions, are now detected out to a red-shift, z$\sim10$, thanks to the James Web Space Telescope (JWST), while the extremely faint recombination lines (RLs) of O$^{++}$ can only be reliably measured in the brightest nearby regions \citep{esteban99, rojas07b, mendez-delgado23, curti23, sanders24, singh26}. 

Gas-phase O$^{++}$ abundances are commonly derived using the so-called ``direct method''. In which, the electron temperature is inferred from ratios of CELs with different excitation energies, while the electron density is obtained independently from ratios of CELs with close-lying upper levels, whose relative populations are set by the density-dependent competition between radiative decay and collisional de-excitation.
Given $T_e$ and $n_e$, the emissivities of O$^{++}$ CELs can be calculated directly, allowing the O$^{++}$/H$^+$ abundance ratio to be determined by comparing their brightness to that of hydrogen RLs \citep{osterbrock06}.

Because CEL emissivities depend exponentially on $T_e$, the direct method is intrinsically sensitive to the thermal structure of a nebula. Hotter regions contribute disproportionately to the observed emission, while cooler regions contribute comparatively little once collisional excitation becomes inefficient. As a result, if present, unresolved temperature variations along the line of sight can bias the inferred $T_e$ toward higher values, leading to systematically underestimated O$\rm ^{++}$ abundances from CELs \citep{peimbert67, mendez-delgado23}.

RLs of elements heavier than Helium (``metals'' in astrophysical lingo) provide an alternative abundance diagnostic that is largely insensitive to temperature fluctuations. Because the emissivities of O$\rm ^{++}$ and H$\rm ^+$ RLs have a very similar temperature dependence, their ratio provides an almost temperature-independent measure of the O$\rm ^{++}$ ionic abundance \citep{osterbrock06, peimbert17}. However, O$\rm ^{++}$ RLs are intrinsically 3 to 4 orders of magnitude fainter than their CEL counterparts, restricting their detection to the brightest and nearest \hii regions.

Since the early work of \cite{wyse42}, it has been recognized that abundances derived from heavy-element RLs systematically exceed those obtained from CELs in \hii regions, often by factors of up to 3  \citep{peimbert93, esteban99, rojas07, rojas07b, esteban09, blanc15, mendez-delgado23, singh26}. This long standing offset is quantified through the abundance discrepancy factor (ADF), defined as the logarithmic difference between RL- and CEL-based abundances. Despite decades of observational and theoretical effort, this discrepancy has persisted as a fundamental obstacle to the accurate determination of chemical abundances in ionized nebulae for more than 80 years.

Resolving this discrepancy is imperative. If heavy-element abundances in star-forming regions are underestimated by a factor of 2--3, the metal content of galaxies and the cosmic heavy-element budget would be systematically biased \citep{peeples14}. Such an offset would propagate across a wide range of astrophysical inferences, directly impacting constraints on physical processes associated with metal production \citep[e.g., nucleosynthesis, ][]{kobayashi20}, ISM enrichment  \citep[heavy element stellar yields, stellar winds and supernovae feedback, ][]{nomoto13, hopkins18}, the baryon cycle on galactic scales  \citep[gas inflows and outflows, ][]{tumlinson17, peroux20}) and heavy element depletion \citep[e.g. dust formation and metal capture in stellar remnants][]{dwek98}). 

The metallicity of the ISM further regulates gas cooling and the resulting star formation, shaping the stellar initial mass function both in theoretical models \citep{hocuk10, dopcke11} and observations \citep{zhou19, liang21}, with further implications for models of the Galactic habitable zone, shifting our expectations of where and when a galaxy like the Milky Way, is most likely to form life-hosting planets like Earth \citep{lineweaver04, johnson12, mejia22}.
Furthermore, since direct-method (i.e. CEL) abundances underpin most of the strong-line diagnostics used to infer chemical abundances in galaxies across cosmic time \citep{blanc15, kewley19, maiolino19, sanders25}, any systematic bias in the local CEL abundance scale would propagate into our understanding of galaxy chemical evolution \citep{moustakas10}.

The physical origin of the abundance discrepancy remains debated. Several hypotheses have been proposed, including shocks and turbulence by stellar feedback \citep{peimbert91, berg15, vaught24}; chemically enriched high density inclusions embedded within the ionized gas \citep{tsamis03, stasinska07}, and departures from a Maxwellian electron energy distribution in the ionized ISM \citep{nicholls12}. 
Many of these scenarios invoke spatial variations in the thermal structure of the plasma, commonly described through the temperature fluctuation formalism of \citep{peimbert67}.

Studies based on slit-spectroscopy of \hii regions have provided evidence supporting a connection between the ADF and temperature fluctuations \citep{peimbert93, rojas07b, mendez-delgado23}, but limit these comparisons to the areas covered by the slits, and do not provide an integral view across the nebula \citep{delgado08}. Spatially resolved observations probing the interior of individual \hii regions are therefore essential to test these claims and to place direct observational constraints on the origin of the ADF.

Using deep, ultra-wide-field integral field spectroscopy from the Sloan Digital Sky Survey V \citep[SDSS-V, ][]{kollmeier26} Local Volume Mapper \citep[LVM, ][]{drory24}, \cite{singh26} mapped the Milky Way \hii region M8, better known as the Lagoon Nebula, at a spatial resolution of $0.21$~pc per spaxel, enabling the first spatially resolved investigation of the O$\rm ^{++}$ abundance discrepancy across an entire \hii region. \cite{singh26} used these data to produce the first spatially resolved IFU maps of O$\rm ^{++}$ RLs and the ADF across an \hii region, revealing that the discrepancy peaks near the main ionizing source, Herschel 36 \citep[Her36, the brightest component of a trinary system O7.5+ star][]{arias10}, and decreases with distance away from it. This behavior is consistent with the slight increase in the ADF reported by \citet{delgado08} in the central $\approx200"$ region of the Orion Nebula using slit-spectroscopy.

In this work, we establish the physical origin of the observed abundance discrepancies observed across the Lagoon Nebula. In Section~\ref{sec:sec2}, we describe the data. Sections~\ref{sec:sec3} outlines the methodology involving measurements of different temperature diagnostics and a simple plasma temperature perturbation model. We present the results in Section~\ref{sec:sec4}, followed by Discussion in Section~\ref{sec:sec5}. We summarize the main conclusions in Section~\ref{sec:sec6}.

\section{Data}
\label{sec:sec2}

We use integral field spectroscopic observations of M8 obtained as part of SDSS-V LVM early science program. The observations and data reduction are described in detail in \cite{singh26}, who presented spatially resolved maps of the physical conditions, O$^{++}$ abundance, and the first spatially resolved map of the ADF, across the entire nebula. Here, we use these measurements to investigate the origin of the abundance discrepancy. To assess radial trends and the effects of spatial averaging, we perform the analysis at three spatial scales: (1) individual spaxel spectra providing fully resolved maps; (2) spectra co-added in $1'$~(0.36 pc) wide annuli centered on Her36; and (3) the average M8 spectrum obtained by averaging over the full FOV. 

\begin{figure}[t]
  \centering
  \includegraphics[width=0.499\textwidth]{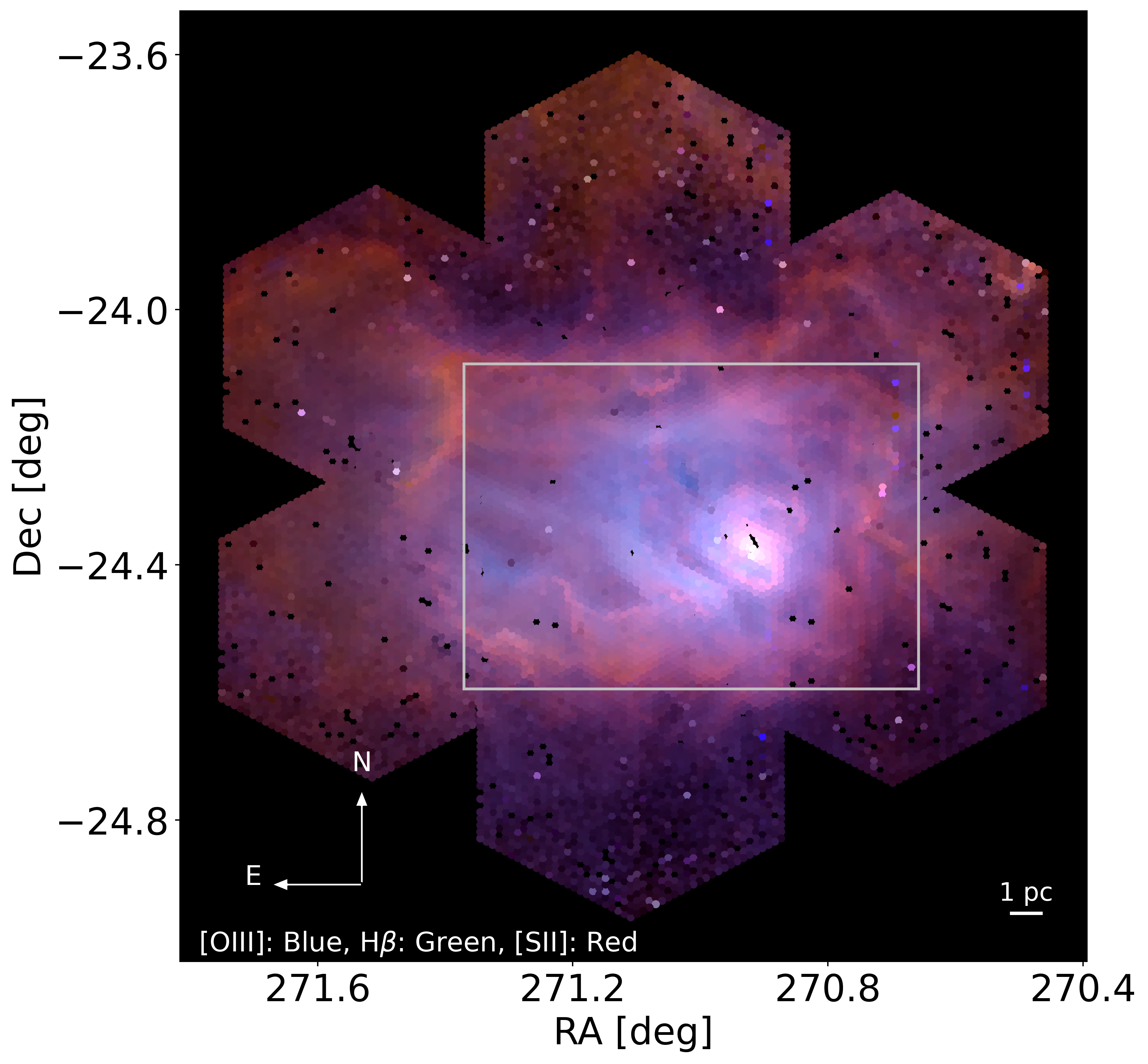}
  \caption{SDSS-V LVM RGB composite image of M8, with \sii$\lambda\lambda$6717+6731 in red, \hb in green, and \oiii$\lambda$5007 in blue; the dark spaxels correspond to masked spaxels. The gray box represents a zoom-in region satisfying the $S/N>3$ for $\lambda 4363$, the subsequent diagnostic maps are shown within this gray box in Section~\ref{sec:sec4}.
  } 
  \label{fig:rgb_composite}
\end{figure}

The LVM observations cover the full extent of the nebula, as shown in Figure \ref{fig:rgb_composite}. However, the spaxels satisfying the signal-to-noise ($S/N$) requirements for the CEL and RL-based diagnostics are concentrated within the central part of the observed field. We therefore define the region enclosed by the gray outline in Figure \ref{fig:rgb_composite} as the FOV shown in the subsequent analysis maps; this region contains all measurements used in this work. For the CEL-based measurements, we require $ S/N>3$ on \oiii$\lambda$4363. For RL-based diagnostics, we use the \orl V1 multiplet, consisting of eight lines: 4638.86, 4641.81, 4649.13, 4650.84, 4661.63, 4673.73, 4676.23, and 4696.36 $\AA$. The two faintest lines, $\lambda$ 4673.73 and $\lambda$4696.36 are not reliably detected in our data, and therefore we use the remaining six lines. For measurements based exclusively on $O^{++}$ RLs, we require $S/N>3$ for the sum of two brightest V1 lines, $\lambda$4649 and $\lambda$4650. For quantities that combine CEL and RL measurements, both the \oiii$\lambda$4363 and \orl V1 criteria are applied simultaneously \citep{singh26}. All strong CELs and HI lines used in this study have a $ S/N>5$.

\section{Methods}
\label{sec:sec3}

\subsection{$\Delta T_e$ and ADF estimation}
\label{sec:sec3.1}

In this study we make use of two independent electron temperatures diagnostics for O$^{++}$ derived in \cite{singh26}. The CEL-based (i.e. direct-method) temperature, $T_e^{CEL}(\rm O^{++})$, is derived from the \oiii$\lambda$4363/\oiii$\lambda$(4959+5007) ratio, and the hybrid temperature, $ T_e^{RLs/CEL}(\rm O^{++})$, is derived from the ratio of the summed V1 RL flux to \oiii$\lambda$5007 CEL. Because the latter diagnostic combines RLs with a CEL, its temperature dependence is different than that of a purely CEL ratio based diagnostic \citep{peimbert13, singh26}. 
Both temperatures are calculated using electron densities from the $S^{+}$ diagnostic based on the [SII]$\lambda\lambda6717,6731$ doublet: $ n_e(S^{+})$.

The O$^{++}/$H$^+$ abundances from CELs and RLs are derived from the relative intensities of $\lambda4959+ 5007$ and the summed \orl V1 multiplet, respectively, with respect to \hb. For both abundance determinations, we adopt $T_e^{CEL}(\rm O^{++})$ and $ n_e(S^{+})$. Using the same physical conditions in both methods ensures that the difference between the resulting abundances reflects the different sensitivities of CELs and RLs, rather than differences introduced by the adopted nebular conditions. In any case, the impact of $T_e$ is negligible on the RL abundance \citep[see ][for details]{singh26}.

The $n_e$, $T_e$, and ionic abundances are calculated using \texttt{PyNeb} \citep{luridiana09}. The atomic data sources adopted are summarized in Table 3 of \citet{singh26}. Uncertainties are propagated via 50 Monte Carlo realizations, after confirming that higher number of realizations do not significantly change the results. In each realization, the measured line fluxes are perturbed within their measurement errors, and all the derived quantities are recalculated. The final measurements and their uncertainties are the mean and standard deviation of the resulting distributions, respectively.

We quantify the temperature discrepancy as the difference between the temperature diagnostics:

\begin{equation}
     \Delta T_e = T_e^{CEL}(\rm O^{++}) - T_e^{RLs/CEL}(\rm O^{++}),
\end{equation}

and the abundance discrepancy factor as:

\begin{equation}
     \mathrm{ADF}(\mathrm{O}^{++}) = \log\!\left(\frac{\mathrm{O}^{++}}{\mathrm{H}^+}\right)_{\!\mathrm{RLs}} - \log\!\left(\frac{\mathrm{O}^{++}}{\mathrm{H}^+}\right)_{\!\mathrm{CEL}}
\end{equation}

\subsection{Plasma temperature perturbation modeling}
\label{sec:sec3.3}

To quantify how electron temperature inhomogeneities along the line of sight bias the observed line fluxes that are used in the different nebular diagnostics used in Section \ref{sec:sec3.1}, we calculate theoretical predictions using a simple temperature perturbation model based on the $t^2$ formalism of \cite{peimbert67}. 

The mean emissivity of a line whose volume-averaged emission arises from a plasma with spatial temperature fluctuations is expanded in a Taylor series around the emission-measure weighted mean temperature $T_0$:

\begin{equation}
\label{equation:taylor_exp}
     \langle \varepsilon \rangle \approx \varepsilon (T_0) \left[1+\frac{T_0^2}{2}\,\frac{\varepsilon"(T_0)}{\varepsilon(T_0)} \,t^2\right]
\end{equation}

where the normalized variance $t^2 = \langle ( T_e -T_0)^{2} \rangle/T_0^{2}$ parametrizes the fractional amplitude of the fluctuations, and the average temperature $T_0$ is a weighted average electron temperature over the volume under consideration (in this case the integral along the line of sight over the solid angle sustained by the areas over which the spectra is integrated), weighted by the densities of the particles involved in the emission mechanism with emissivity $\varepsilon$. As the CEL emissivity depends exponentially on $ T_e$, hotter regions contribute disproportionately to the observed \oiii$\lambda$4363/$\lambda$(4959+5007) ratio, biasing $T_e^{CEL}(\rm O^{++})$ systematically above $T_0$. The hybrid RL/CEL ratio has a weaker effective temperature sensitivity, so $T_e^{RLs/CEL}(\rm O^{++})$ is less susceptible to the temperature fluctuations in the gas than $T_e^{CEL}(\rm O^{++})$(see Figure~\ref{fig:teo2_t0}).

The emissivities required to evaluate Equation~\ref{equation:taylor_exp} are given by \citet{peimbert13}:

\begin{equation}
\label{equation:em_v1}
    \varepsilon_{V1}(T_e) = C_{V1}\, T_e^{-0.85}, 
\end{equation}

\begin{equation}
    \varepsilon_{4959}(T_e) = C_{4959}\,T_e^{-0.34}\exp\left(-\frac{29160}{T_e}\right)
\end{equation}

\begin{equation}
    \varepsilon_{4363}(T_e) = C_{4363}\,T_e^{-0.12}\exp\left(-\frac{61170}{T_e}\right)
\end{equation}

\begin{equation}
\label{equation:em_hb}
    \varepsilon_{H\beta}(T_e) = C_{H\beta} \,T_e^{-0.86}
\end{equation}

Where the normalization constant $C_{\lambda}$ depends only on atomic parameters, including collisional excitation and de-excitation parameters, radiative transition probabilities, and recombination coefficients. For the $\lambda$4959 transition the constant $C_{4959}=1.74\times10^{-17} erg\, s^{-1} \, cm^{3}$ is derived from the collisional excitation equilibrium condition,

\begin{equation}
    \varepsilon_{4959} = h \nu_{4959} A_{ul} \frac{n_u}{n(\mathrm{O^{++}})},
\end{equation} 

where $n_u$ is the population density of $O^{++}$ in the upper energy level of the $\lambda4959$ transition, $n(O^{++})$ is the total ion density, and $A_{ul}$ is the Einstein spontaneous transition probability for transition from the upper (u) to lower level (l). The collision strengths and transition probabilities from \cite{lennon_burke94} and \cite{osterbrock06}. The ratio $C_{V1}/C_{4959}=6.56\times10^{-5}$ is taken from \cite{peimbert13}, while $C_{H\beta}=2.71\times10^{-23}$ from \cite{osterbrock06}. The \oiii$\lambda$4959/$\lambda$5007 ratio is a constant under typical \hii regions conditions, and both these lines share the same $ T_e$ dependence. All line emissivities are therefore evaluated at $T_e^{CEL}$, which is calculated using equation~\ref{equation:tedm}.  

Applying Equation~\ref{equation:taylor_exp} to each line, \cite{peimbert13} shows that the observed CEL and hybrid temperatures relate to $T_0$ and $t^2$ through:

\begin{equation}
\label{equation:tedm}
     T_e^{CEL}(\rm O^{++}) = T_0(\mathrm{O}^{++}) \left[
    1 + \left(\frac{91300}{T_0(\mathrm{O}^{++})} - 2.68\right)
    \frac{t^2(\mathrm{O}^{++})}{2}
\right]
\end{equation}

\begin{multline}
\label{equation:tehybrid}
T_e^{RLs/CEL}(\rm O^{++})
= T_0(\mathrm{O}^{++})
\Bigg[
1 +
\Bigg(
\frac{29160}{T_0(\mathrm{O}^{++})}
- 3.095 \\
\qquad
+ \frac{0.415}
{\frac{29160}{T_0(\mathrm{O}^{++})}+0.415}
\Bigg)
\frac{t^2(\mathrm{O}^{++})}{2}
\Bigg]
\end{multline}

where $T_0(\rm O^{++})$ is the average electron temperature weighted by the $n_e$ and the ion number density of $\rm O^{++}$.

The ionic abundances derived from each line family are:

\begin{equation}
     \left(\frac{\mathrm{O}^{++}}{\mathrm{H}^+}\right)_{\!\mathrm{CEL}} = \frac{I_{4959}}{I_{H\beta}}\,\frac{\varepsilon_{H\beta}(T_e^{\mathrm{CEL}})}{\varepsilon_{4959}(T_e^{\mathrm{CEL}})},
\end{equation}

\begin{equation}
\label{equation:ab_rl}
     \left(\frac{\mathrm{O}^{++}}{\mathrm{H}^+}\right)_{\!\mathrm{RLs}} = \frac{I_{V1}}{I_{H\beta}}\,\frac{\varepsilon_{H\beta}(T_e^{\mathrm{CEL}})}{\varepsilon_{V1}(T_e^{\mathrm{CEL}})}.
\end{equation}

Where $I_{4959}$ and $I_{H\beta}$ are the intensities of the \oiii$\lambda$4959 and $H\beta$ lines, respectively; and $I_{V1}$ is the sum of individual OII V1 intensities.

Note the use of $T_e^{CEL}$ in equation \ref{equation:ab_rl}. As discussed in Section \ref{sec:sec3.1}, the impact of adopting either $T_e^{CEL}$ or $T_e^{RLs/CEL}$ when calculating the RL abundance is negligible. This is because the exponents in equations \ref{equation:em_v1} and \ref{equation:em_hb} are almost identical.

Equations~\ref{equation:tedm} and \ref{equation:tehybrid} together constrain both $T_0$ and $t^2$ from the measured pair ($T_e^{CEL}$, $T_e^{RLs/CEL}$). We construct a model grid spanning $T_0 = 5000-12000$ K, $t^2=0-0.1$, and intrinsic abundance $12+\log{(\rm O^{++}/H^+})=6.5 - 9.5$ in steps of 0.5 dex, and utilize the above equations to determine the predicted outcome of $T_e^{CEL}(\rm O^{++})$, $T_e^{RLs/CEL}(\rm O^{++})$, CEL and RL-based O$^{++}$ ionic abundances, and finally $ \Delta T_e$ and ADF at each grid point. Resulting figures from Figure~\ref{fig:teo2_t0} to Figure~\ref{fig:delte_adf_t2}, are analyzed in Section~\ref{sec:sec4}.

\section{Results}
\label{sec:sec4}

\subsection{Plasma temperature perturbation model predictions} 


\begin{figure}[t]
  \centering
  \includegraphics[width=0.499\textwidth]{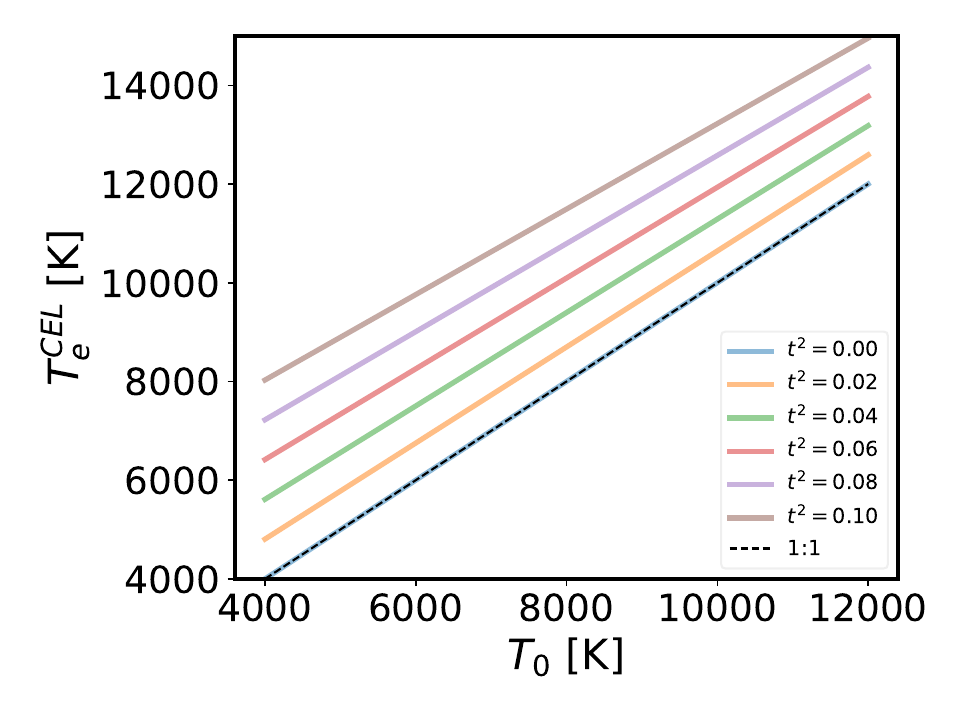}
  \includegraphics[width=0.499\textwidth]{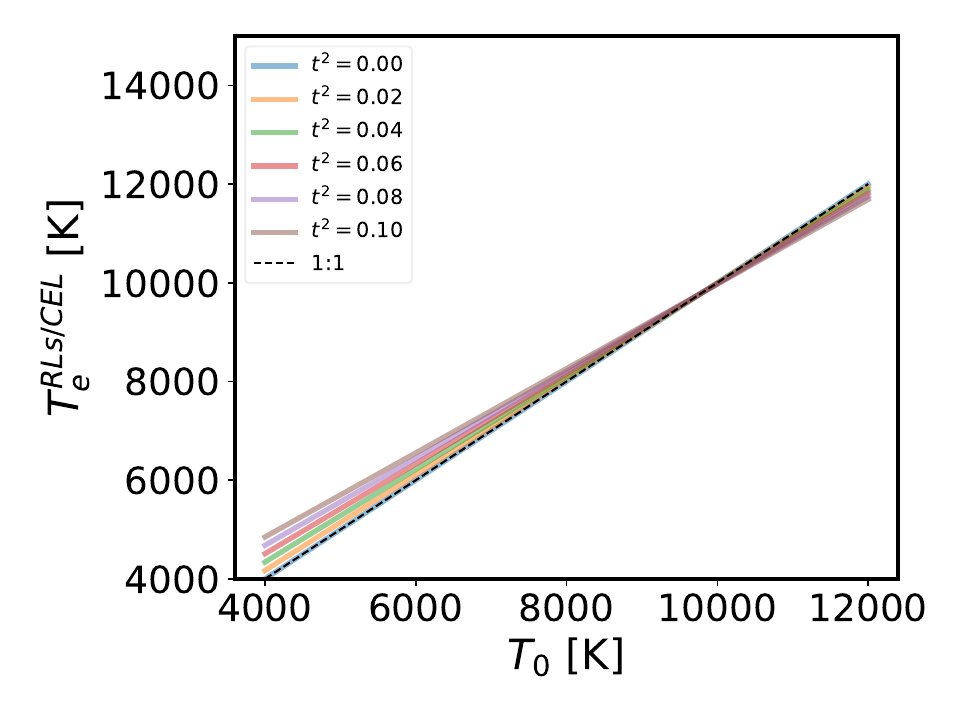}
  \caption{Comparison of measured electron temperatures derived from (a) CELs and (b) RLs/CEL against the intrinsic average electron temperature of the gas, $T_0$ for varying $t^2$. The dashed black line marks the $1{:}1$ relation. At $t^2 = 0$, they follow the intrinsic average temperature of the gas and overlap with the $1{:}1$ line.}
  \label{fig:teo2_t0}
\end{figure}


\begin{figure}[t]
  \centering
  \includegraphics[width=0.499\textwidth]{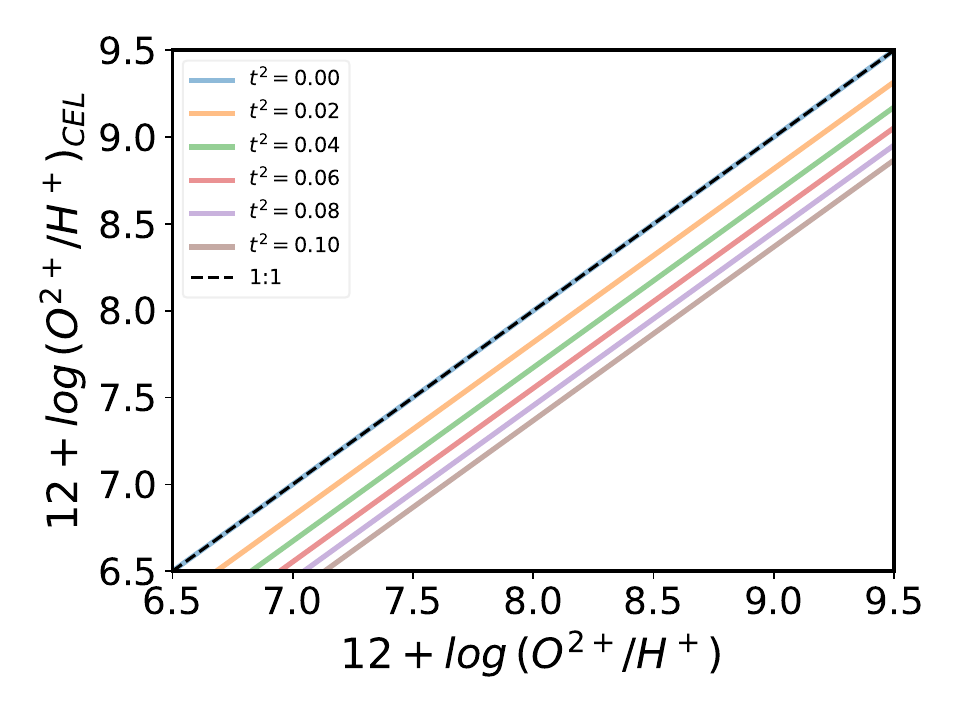}
  \includegraphics[width=0.499\textwidth]{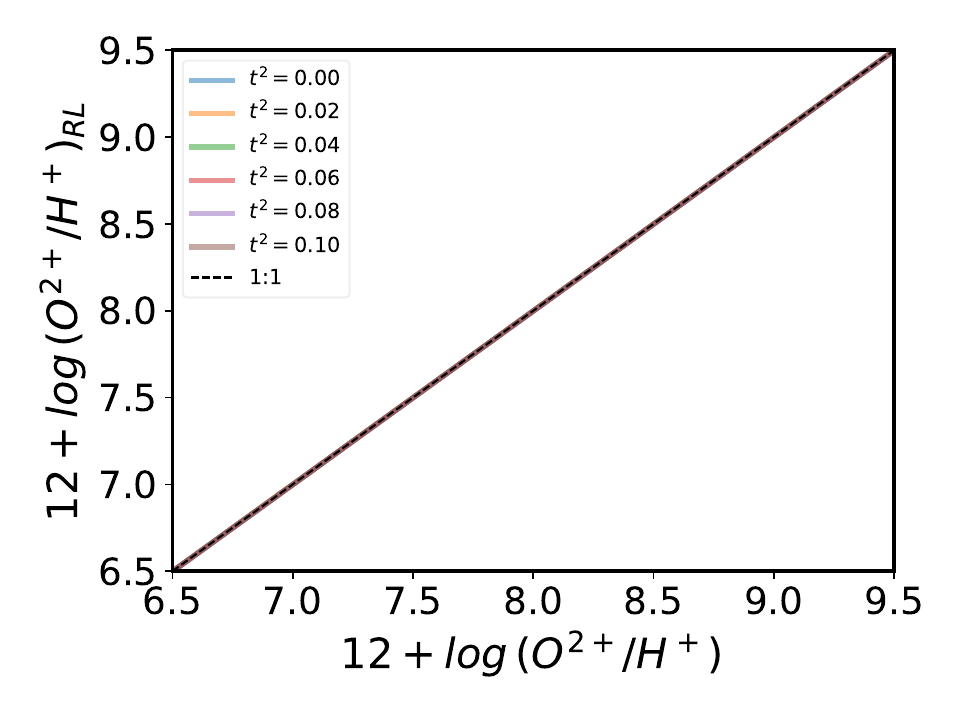}
  \caption{Comparison of $O^{2+}$ ionic abundances derived from CELs and RLs as a function of $t^2$.
  (a) CEL-based $O^{2+}$ ionic abundances are significantly suppressed with increasing $t^2$. 
  (b) RL-based $O^{2+}$ abundances show negligible variation with $t^2$. The dashed black line marks the $1{:}1$ relation. At $t^2 = 0$, they follow the intrinsic abundance of the gas and overlap with the $1{:}1$ line. The ``true abundance'' assumed in the theoretical model is denoted on x-axes of the two plots.}
  \label{fig:opp_cel_rl_opp0}
\end{figure}


\begin{figure}[t]
  \centering
  \includegraphics[width=0.499\textwidth]{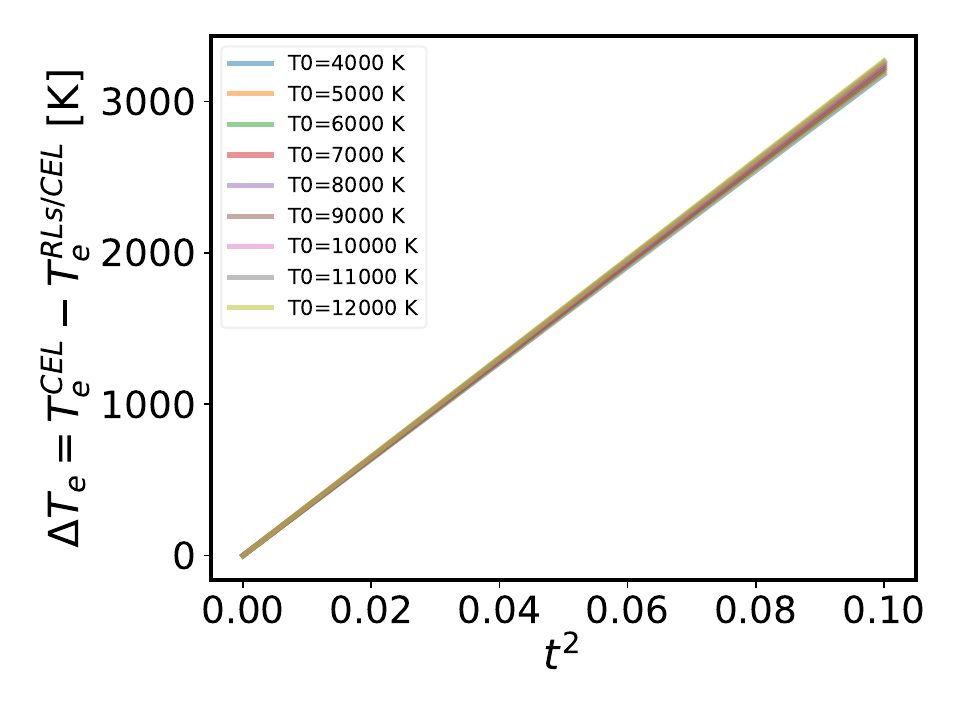}
  \includegraphics[width=0.499\textwidth]{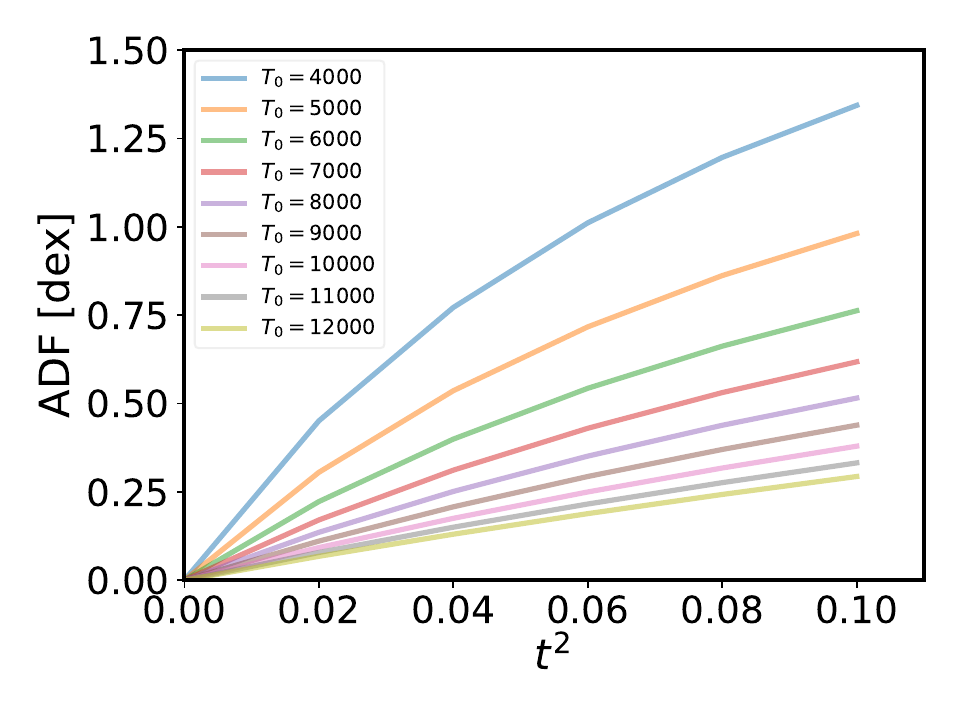}
  \caption{Relation between theoretical temperature fluctuation, temperature discrepancy, and the abundance discrepancy.
  (a) Temperature discrepancy $ \Delta T_e$ as a function of $t^2$, shown for a range of average temperatures $T_0$. The linear trend illustrates how $ \Delta T_e$ traces the amplitude of small-scale temperature fluctuations.
  (b) Abundance discrepancy factor (ADF) as a function of $t^2$, highlighting how the difference between CEL- and RL-based temperatures provides a direct physical link to the observed abundance discrepancies in ionized nebulae.}
  \label{fig:delte_adf_t2}
\end{figure}

The upper panel of Figure~\ref{fig:teo2_t0} shows that temperature fluctuations preferentially bias $T_e^{CEL}(\rm O^{++})$ toward higher values. Because CEL emissivities depend exponentially on the temperature, hotter regions contribute disproportionately to the \oiii$\lambda$4363 emission relative to the \oiii$\lambda$4959,5007 lines when averaging along the line of sight, biasing the \oiii$\lambda$4363/$\lambda$4959 ratio towards high values. In contrast, $T_e^{RLs/CEL}(\rm O^{++})$ is much less sensitive to local temperature inhomogeneities (lower panel) because it is derived from the ratio of RL fluxes to the CEL flux rather than solely on CELs. Figure~\ref{fig:opp_cel_rl_opp0} shows that thermal fluctuations induce a systematic underestimation of CEL-based $O^{++}$ abundance (mainly through their impact on the derivation of $T_e$) while they leave the RL-based $O^{++}$ abundance largely unaffected. 

Importantly, Figure~\ref{fig:delte_adf_t2} show that the $\Delta T_e$ increases linearly with $t^2$, with only a weak dependence on $T_0$. This is a notable aspect of this work, as we have identified an observational quantity ($\Delta T_e$), which by itself provides an excellent observational proxy of the electron temperature fluctuations amplitude in the ionized ISM. 

As it has been long established in the literature, models predicts that the ADF increases with $t^2$. The relation is not linear, as it flattens for high $t^2$ values, and its normalization is highly dependent on the value of the average underlying electron temperature $T_0$ (see Figure~\ref{fig:delte_adf_t2}). Previous ADF studies have simply used this relation to infer a particular value of $t^2$ to reproduce an observed ADF. The novel aspect of this work is the introduction of the observable quantity $ \Delta T_e$, which, when correlated with the ADF, allows us to test directly whether the observed behavior of a real \hii region on spatially resolved scales is consistent with that expected in the presence of temperature fluctuations.

\subsection{Spatially resolved $\Delta T_e$ and ADF} 

\begin{figure}[t]
  \centering
  \includegraphics[width=0.499\textwidth]{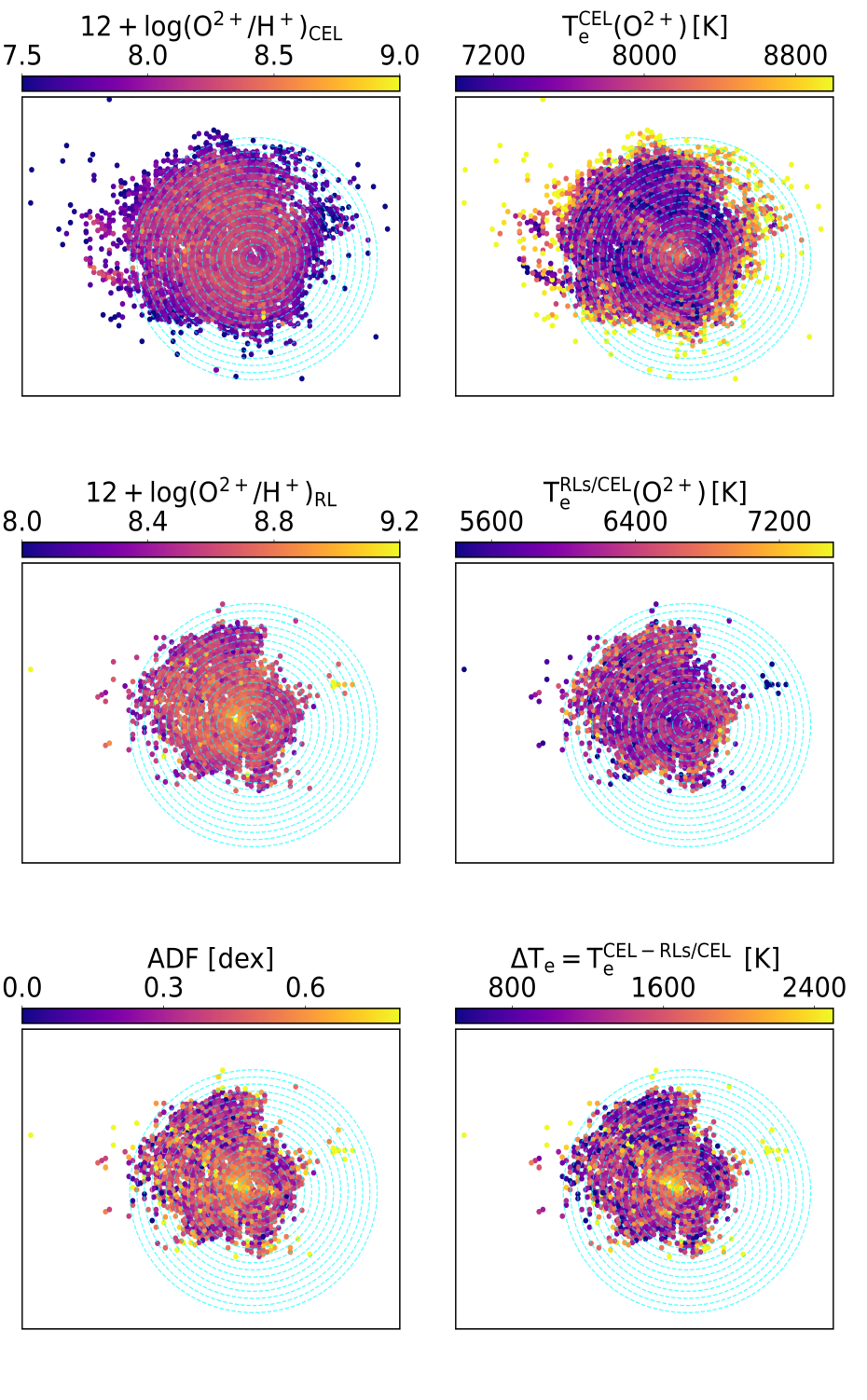}
  \caption{Spatially resolved maps, within the zoomed region and overlaid with the spatial binning scheme in light cyan, presented in this work -- $ 1'$ wide annular bins out to $ 17'$, where the faint O$^{++}$ RLs are still reliable. \textit{Left column:} Spatially resolved maps of measured CEL- and RL-based O$\rm ^{++}$ abundances and the ADF(O$\rm ^{++}$); \textit{Right column:} Spatially resolved maps of CEL, RLs/CEL based $ T_e (\rm O^{++}$), and $ \Delta T_e (\rm O^{++}$).
  } 
  \label{fig:diagnostics_maps}
\end{figure}

We examine the spatially resolved, $1'$ ($0.36$~pc) wide annular-binned, and average measurements of $T_e$ and the O$\rm ^{++}$ abundance across the M8 nebula. This allows us to characterize the spatial structure of the nebula and assess how the observed quantities relate to each other and change with spatial averaging.

Figure~\ref{fig:diagnostics_maps} presents spatially resolved maps of the key quantities in this analysis. The RL-based O$\rm ^{++}$ abundances are systematically higher than those derived from CELs across the entire nebula. Their logarithmic difference (the ADF) shows a clear spatial trend, with higher values near the ionizing center and declining outward. 

The two temperature diagnostics show markedly different spatial behavior. The CELs-based electron temperature $T_e^{CEL}(\rm O^{++})$ increases toward the ionization center. As will be shown below, this behavior is most likely an artifact associated with enhanced temperature fluctuations, and their effect on the luminosity weighting of CELs toward the central region. In contrast, the hybrid electron temperature $T_e^{RLs/CEL}(\rm O^{++})$, which combines RLs and CELs, is systematically lower than the CEL-based $T_e$ and remains spatially uniform across the nebula. 

Since all the diagnostics are based on O$^{++}$ emission, these differences can not be attributed to the different ionic volumes traced by different ions. It is instead a signal that the gas near Her36 is more thermally inhomogeneous \citep{singh26}.

The temperatures difference, $\Delta T_e(\rm O^{++})$, peaks in the central regions and decreases outward, closely tracing the spatial profile of the ADF, suggesting a direct connection between the temperature discrepancy and the abundance discrepancy. 


\begin{figure*}[t]
  \centering
  \includegraphics[width=\textwidth]{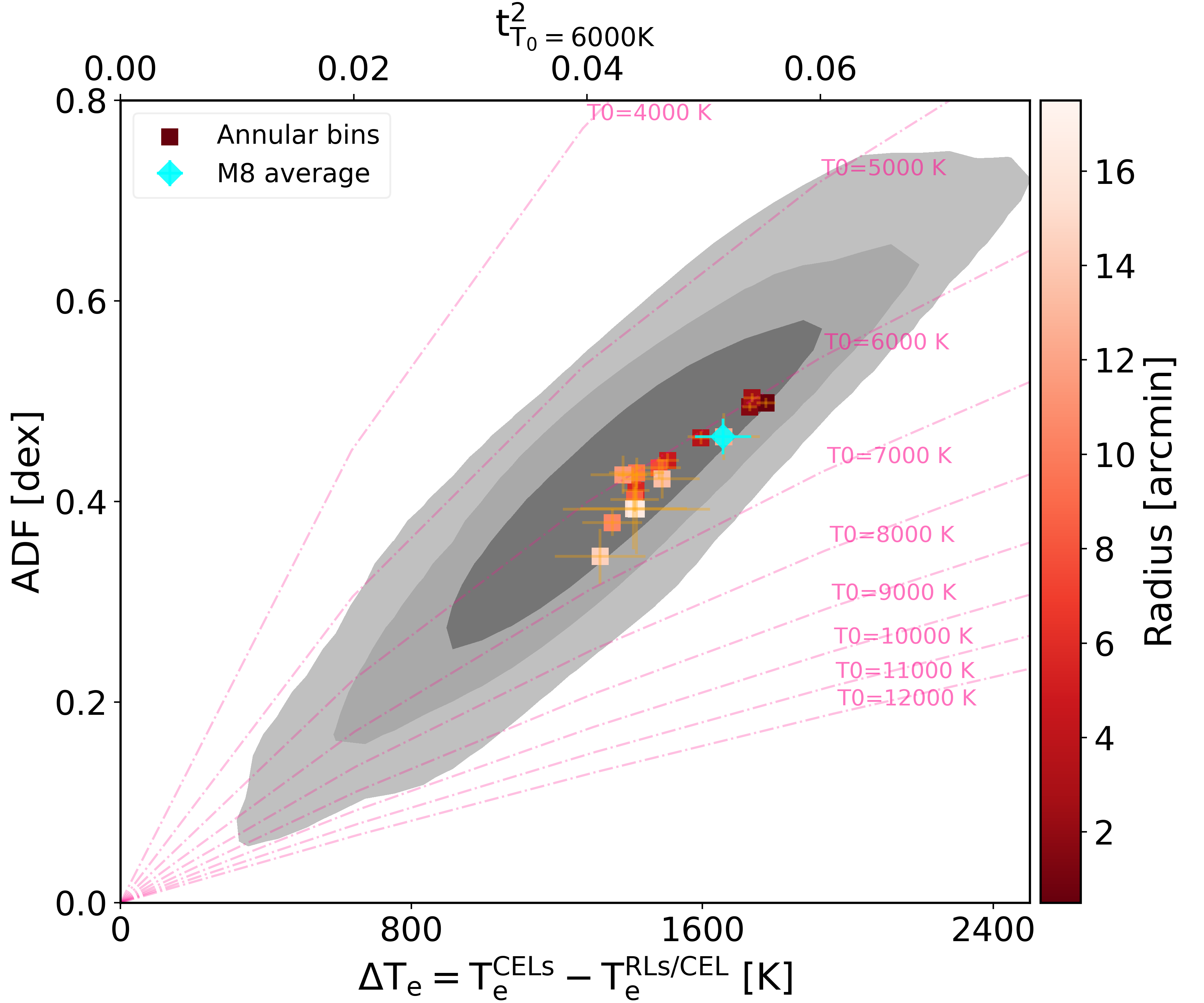}
  \caption{The ADF is shown as a function of $ \Delta T_e$. Gray density contours represent spatially resolved measurements. Colored squares denote annular binned measurements, color-coded by projected distance from the Her36, while the cyan diamond indicates the average measurements over the full M8 field. 
  Pink curves show theoretical predictions for different average temperatures of the nebula, $T_0$, with the upper axis indicating the corresponding temperature fluctuation parameter, $t^2$. The observed measurements closely follow the theoretical average temperature, $ T_0 = 6000$K.
  }
  \label{fig:adf_del_te_correlation}
\end{figure*}


\subsection{The $\Delta T_e$--ADF correlation}

Figure~\ref{fig:adf_del_te_correlation} presents the ADF measured in each annuli as a function of the temperature discrepancy ($ \Delta T_e$). The spatially resolved measurements are also shown as density contours encompassing 50\%, 75\%, and 90\% of the measurements. Regions with higher levels of abundance discrepancy between RL and CEL diagnostics, also show higher levels of electron temperature discrepancies. 

The observed ADF and $ \Delta T_e$, are highly correlated across every scale we probe. The spatially resolved measurements yield a Pearson correlation coefficient of 0.93 (p-value=$ 10^{-6}$), confirming a robust point-to-point correspondence throughout the full FOV. The $1'$ ($0.36$~pc) wide annular binned measurements yield the same correlation coefficient of 0.93 (p-value=$ 4.4\times10^{-8}$). Measurements from the average spectrum of the entire M8 lie along the same relation \citep{singh26} demonstrating that the correspondence between ADF and $ \Delta T_e$ is preserved across all spatial scales probed.

\subsection{Temperature-fluctuation and the $\Delta T_e$ relation} 

The observed relation can be directly compared with the predictions of the temperature fluctuation model described in Section~\ref{sec:sec3.3}. Figure~\ref{fig:adf_del_te_correlation} presents the theoretical predictions (pink curves) for the way in which the ADF and $ \Delta T_e$ quantities should relate to each other in the presence of electron temperature fluctuations. 

The upper axis of Figure~\ref{fig:adf_del_te_correlation} provides the mapping between $ \Delta T_e$ and $t^2$ (at a fixed $T_0=6000$, although the effect of this parameter is very small).
Comparing the data in Figure~\ref{fig:adf_del_te_correlation} to the model predictions, we observe that the M8 region behaves as expected for a plasma at roughly constant average electron temperature, in the presence of electron temperature fluctuations whose amplitude decreases as one moves away from the central ionizing source. 

Although the hybrid $T_e^{RLs/CEL}(\rm O^{++})$ is substantially a better electron temperature diagnostic than $T_e^{CEL}$ alone, it remains subject to measurable biases (see Figure \ref{fig:teo2_t0}). By jointly modeling the impact of $T_e$ fluctuations on both CELs and RLs simultaneously, we can constrain $T_0$ more accurately than either diagnostic in isolation. This approach provides a quantitative constraint on both the average electron temperature, $T_0$, and the amplitude of the temperature fluctuations, $t^2$.

\subsection{Resolving the abundance discrepancy} 

The average M8 spectrum yields a best-fit average electron temperature of $T_0=6240^{+310}_{-360}$~K. This value is consistent, within the uncertainties, with the hybrid $T_e^{RLs/CEL}(\rm O^{++}) = 6364\pm49$, but is $\sim$ 1800~K lower than the CEL-based $T_e^{CEL}(\rm O^{++}) = 8025\pm67$ K \citep{singh26}. The annular binned spectra recover a similar characteristic temperature, with a value of $T_0=6020^{+30}_{-20}$~K, obtained by fitting all the points assuming a constant temperature model. While this a good representative estimate for the nebula, close inspection of Figure \ref{fig:adf_del_te_correlation} does hint for a small increase of a few hundred K towards the outermost bins. The amplitude of the root-mean-square temperature fluctuations inferred from the average M8 spectrum is $22\pm 1$\% ($t^2=0.050$), decreasing from a central value of $24\pm 0.4$\% ($t^2=0.056$) down to $20\pm 3$\% ($t^2=0.043$) for the outermost annular bin. These values are consistent with $t^2=0.045$ reported by \citet{rojas07} for a localized region near the center of M8.

Adopting a value of $ T_0=6240$~K to derive both RL-and CEL-based O$^{++}$ abundances from the average spectrum of the nebula, reduces the abundance discrepancy in M8 from $0.47\pm0.02$ dex to $-0.04\pm0.02$ dex, effectively solving the abundance discrepancy problem. These findings establish that temperature inhomogeneities as the dominant origin of the abundance discrepancy in M8. They further demonstrate that jointly modeling the effects of temperature fluctuations across multiple diagnostics is the most reliable path to obtain unbiased temperature and abundance measurements.



\section{Discussion}
\label{sec:sec5}

Despite decades of investigation, identifying a direct observational connection between the CEL vs RL abundance discrepancy and its physical origin in \hii regions has remained challenging, especially because previous observations have generally lacked sufficient spatial coverage and resolution required to resolve the internal structure of individual nebulae \citep[][among others]{esteban99, rojas07, peimbert13}. \citet{delgado08} and \citet{delgado10} investigated spatial variations in ADF($O^{++}$) and their relation to the thermal structure using long-slit spectroscopy. \citet{delgado08} reported a weak increase in the ADF toward the ionizing center of Orion, while \citet{delgado10} observed the central $\approx200"$ region of M8 and found a largely homogeneous ADF distribution. These studies provide important spatial constraints, although the slit-based coverage limited the comparison to the areas sampled by the slit rather than providing an integral view of the nebula.

Using the SDSS-V Local Volume Mapper spatially resolved observations of M8, \cite{singh26} argued that the abundance discrepancy in this nebula is not driven by density variations or by the presence of distinct chemically inhomogeneous phases \citep[unlike scenarios proposed for PNe,][]{tsamis08, rojas16, ali&dopita19, rojas22}. The authors suggest that $T_e$ inhomogeneities could play a central role in producing the discrepancy. 

The interpretation of the abundance discrepancy in terms of $T_e$-fluctuations has been debated for decades \citep{esteban99, rojas07, delgado08, mendez-delgado23, yuguang23}. In particular, the role of $T_e$ fluctuations in biasing direct-method CEL-based chemical abundances has recently been put into question. \citet{yuguang23} tested the temperature fluctuation hypotheses in Mrk\,71 using far-infrared CELs of O$^{++}$, finding their infrared abundance consistent with the optical CEL abundance rather than the RL abundance, which they interpreted as ruling out temperature fluctuation in that object. This conclusion has been contested by \citet{mendez-delgado24}, who show that it is sensitive to the assumed flux calibration, reddening correction, and electron density, and that alternative treatment of the same data are consistent with the $t^2$ formalism. 

Our spatially resolved analysis of the LVM data on M8 establishes that the sub-parsec spatially-resolved behavior of the nebular emission inside this HII region, is fully consistent with the temperature fluctuations picture. 
We present the most direct observational evidence to date for $T_e$ inhomogeneities being the primary driver of the CEL vs RL abundance discrepancy in \hii regions. This highlights the value of sub-pc integral-field spectroscopic observations of ionized nebulae, to resolve the micro-physics of interstellar ionized gas emission mechanisms. 
 
The quantity $\Delta T_e$, the difference between direct method and hybrid temperature diagnostics, which can be measured directly from deep HII region spectra, offers an excellent observational proxy for $t^2$. The methodology presented here, in which $\Delta T_e$ is correlated with the observed ADFs across different areas inside an \hii region, provides a new method to directly constrained not only $t^2$, but most importantly $T_0$, the average electron temperature of the emitting ionized gas. This is important because the ADF does not uniquely constrain the thermal structure: its relation to $t^2$ depends on $T_0$. Our results demonstrate that jointly considering ADF and $\Delta T_e$ provides a more direct constraint on the underlying thermal structure, which can be used to reliably recover an abundance scale consistent between CELs and RLs.

As wide-field integral field spectroscopy with facilities such as SDSS-V LVM extends resolved coverage to larger samples of \hii regions across a range of ionizing conditions and metallicities, $ \Delta T_e$ measurements can provide the statistical calibration of the $t^2$ correction the field has lacked, moving the Peimbert correction \citep{peimbert67} from a case-by-case assumption to an empirically grounded tool that could be applied generally. 

A broader consequence of our result concerns the overall metal budget of galaxies, known as the ``missing metals problem'', which arises from the discrepancy between the cosmic density of metals expected to have been produced by previous generations of star formation and that accounted for in stars, the ISM, the circumgalactic medium (CGM), and intergalactic medium (IGM). Current estimates suggesting 35-90 $\%$ of metals remain unaccounted for \citep{ferrara05, bouche07, pagel08}. Surveys such as COS-Halos \citep{peeples14} find that only about $20-25\%$ of the metals produced over a galaxy's lifetime are retained in stars, ISM gas, and dust, with the remainder presumed to reside in the CGM and IGM. The ISM component of this census rests on gas-phase metallicities derived from CELs. If temperature fluctuations of the kind characterized here cause CEL-based methods to systematically underestimate the true ISM oxygen-abundance, then the metal budget of galaxies would likewise be underestimated. Accounting for these effects may partially alleviate this missing metal/oxygen problem. The consequences of resolving the abundance discrepancy extend from individual galactic \hii regions to the cosmological accounting of metal production and retention in galaxies.

Along the same lines, a direct implications of this work concerns extragalactic metallicity measurements. The JWST has enabled auroral line detections at redshifts approaching z$\sim$10, allowing the direct method to be applied to galaxies across all cosmic times, and the empirical calibration of strong-line metallicity diagnostics that are applicable at different redshifts \citep{sanders24, laseter24, sanders25, scholte25}. These galaxies are, however, entirely unresolved; their integrated spectra sum together numerous \hii regions spanning a range of ionization parameters, radiation field hardnesses, densities, and temperatures, none of which are individually separable from the integrated light \citep{sanders18}. In principle, the resulting metallicities from these methods trace the luminosity weighted metallicity of the individual \hii regions. But the biases associated with the thermal structure of these regions are not accounted for and should affect these diagnostics, biasing our estimates of the early chemical evolution of the universe.

The thermal structure of \hii regions depends on the clumpiness of the neutral ISM, the hardness of the ionizing radiation field, the gas metallicity, and the local pressure, all of which evolve substantially with redshift. While the magnitudes of the temperature fluctuations we measure in M8 may differ substantially from those in galaxies at cosmic noon or in the epoch of reionization, their impact on the emission mechanisms constrained here is the same one that is currently ignored in galaxy-scale CEL-based metallicity measurements. Establishing how the thermal structure of HII regions affects chemical abundance diagnostics in a resolved nearby systems such as M8, is a necessary first step before corrections for these effects can be properly propagated into high-z chemical abundance diagnostic calibrations.

\section{Conclusion}
\label{sec:sec6}

Using spatially resolved SDSS-V Local Volume Mapper observations of M8, we investigate the origin of the long-standing abundance discrepancy between CELs and RL. Our main conclusions are:

\begin{enumerate}

    \item We find a tight spatial correlation between the $O^{++}$ ADF and $\Delta T_e$. The agreement of this observed relation with the theoretical predictions of the temperature-perturbation formalism, shows that $T_e$ fluctuations are the primary drivers of the CEL--RL abundance discrepancy in M8. The electron temperatures inferred from the direct method are biased toward higher temperatures, leading to a systematic underestimation of the oxygen abundance when temperature fluctuations are ignored.

    \item The observed $\Delta T_e$--ADF relation, together with the temperature-perturbation model, provides an empirical way to constrain the thermal structure of an \hii region. In particular, $\Delta T_e$ increases linearly with the square of the amplitude of the temperature fluctuations ($t^2$), with only a weak dependence on the underlying average electron temperature ($T_0$). It therefore serves as a direct observational proxy of $t^2$ and a means of constraining the thermal structure of the nebula.

    \item The mean temperature, $T_0$, constrained from the $\Delta T_e$--ADF relation, reconciles the CEL- and RL- based $\rm O^{++}$ abundances in M8, effectively solving the abundance discrepancy problem. This demonstrates that accounting for the thermal structure of the plasma in ionized nebulae is essential for obtaining unbiased CEL-based abundances.

\end{enumerate}

The broader significance of these results is that they establish the thermal structure of \hii regions as an important factor in assessing the reliability of nebular abundance measurements. Determining whether the temperature fluctuations identified in M8 are common across different ISM environments is therefore a critical next step. The After-Sloan 5 (AS5) Local Volume Mapper 2 (LVM2) survey will extend this analysis to hundreds of \hii regions across the Milky Way and nearby galaxies, providing the faint-line measurements needed to systematically measure temperature fluctuations and RL-based abundances across a range of physical conditions and metallicities. M8 thus provides a resolved benchmark for testing how the thermal structure of ionized gas shapes abundance measurements across the local Universe.

\section*{Acknowledgments}

A.S. acknowledges the support from the ANID Basal projects FB21250718. G.A.B. and E.J. acknowledge the support from the ANID Basal project FB210003. C.M. acknowledges the support of grant UNAM/DGAPA/PAPIIT IG101224 and IN113226. J.E. M-D. and C. M. thank the support by SECIHTI CBF-2025-I-2048 project ``Resolviendo la Física Interna de las Galaxias: De las Escalas Locales a la Estructura Global con el SDSS-V Local Volume Mapper'' (PI: M\'endez Delgado). 
KK and EE gratefully acknowledge funding from the European Research Council’s starting grant ERC StG-101077573 (“ISM-METALS"). OE acknowledges funding from the Deutsche Forschungsgemeinschaft (DFG, German Research Foundation) -- project-ID 541068876. J.G.F-T gratefully acknowledges the support provided by ANID Fondecyt Regular No. 1260371.

Funding for the Sloan Digital Sky Survey V has been provided by the Alfred P. Sloan Foundation, the Heising-Simons Foundation, the National Science Foundation, and the Participating Institutions. SDSS acknowledges support and resources from the Center for High-Performance Computing at the University of Utah. SDSS telescopes are located at Apache Point Observatory, funded by the Astrophysical Research Consortium and operated by New Mexico State University, and at Las Campanas Observatory, operated by the Carnegie Institution for Science. The SDSS web site is \url{www.sdss.org}.

SDSS is managed by the Astrophysical Research Consortium for the Participating Institutions of the SDSS Collaboration, including the Carnegie Institution for Science, Chilean National Time Allocation Committee (CNTAC) ratified researchers, Caltech, the Gotham Participation Group, Harvard University, Heidelberg University, The Flatiron Institute, The Johns Hopkins University, L'Ecole polytechnique f\'{e}d\'{e}rale de Lausanne (EPFL), Leibniz-Institut f\"{u}r Astrophysik Potsdam (AIP), Max-Planck-Institut f\"{u}r Astronomie (MPIA Heidelberg), Max-Planck-Institut f\"{u}r Extraterrestrische Physik (MPE), Nanjing University, National Astronomical Observatories of China (NAOC), New Mexico State University, The Ohio State University, Pennsylvania State University, Smithsonian Astrophysical Observatory, Space Telescope Science Institute (STScI), the Stellar Astrophysics Participation Group, Universidad Nacional Aut\'{o}noma de M\'{e}xico, University of Arizona, University of Colorado Boulder, University of Illinois at Urbana-Champaign, University of Toronto, University of Utah, University of Virginia, Yale University, and Yunnan University.

\facilities{SDSS-V Local Volume Mapper}

\software{astropy \citep{2013A&A...558A..33A,2018AJ....156..123A,2022ApJ...935..167A},  
        Scipy \citep{virtanen20}, Numpy \citep{harris20}, Matplotlib \citep{hunter07},          }

\bibliography{references}

\end{document}